\documentclass[preprint]{revtex4}
\usepackage{graphicx}
\usepackage{subfigure}
\usepackage{multirow}
\usepackage{hyperref}
\usepackage{amsmath,bm,amssymb,amsfonts}
\usepackage{color}
\newcommand{\dalm}{\kern1pt\vbox{\hrule height 0.9pt\hbox{\vrule width 0.9pt
			\hskip 2.5pt\vbox{\vskip 5.5pt}\hskip 3pt\vrule width 0.3pt}\hrule height 0.3pt}
	\kern1pt}
\begin{document}
	\title{{\bf
			Scalarized Hybrid Neutron Stars in Scalar Tensor Gravity}}
	
	\author{{\bf Fahimeh Rahimi $^{1,2}$ } and {\bf Zeinab Rezaei $^{1,2}$} \footnote{Corresponding author. E-mail:
				zrezaei@shirazu.ac.ir}}
	\affiliation{ $^{1}$Department of Physics, School of Science, Shiraz
University, Shiraz 71454, Iran.\\
		$^{2}$Biruni Observatory, School of Science, Shiraz
University, Shiraz 71454, Iran.}
	
	
	\begin{abstract}
Hybrid neutron stars, the compact objects consisting hadronic matter and strange quark matter, can be considered as the probes for the scalar tensor gravity.
In this work, we explore the scalarization of hybrid neutron stars in the scalar tensor gravity. For the hadronic phase, we apply a piecewise polytropic equation of state constrained by the observational data of GW170817 and the data of six low-mass X-ray binaries with thermonuclear burst or the symmetry energy of the nuclear interaction. In addition, to describe the strange quark matter inside the hybrid neutron star, different MIT bag models are employed. We study the effects of the value of bag constant, the mass of s quark, the perturbative quantum chromodynamics correction parameter, and the density jump at the surface
of quark-hadronic phase transition on the scalarization of hybrid neutron stars. Our results confirm that the scalarization is more sensitive to the value of bag constant, the mass of s quark, and the density jump compared to the perturbative quantum chromodynamics correction parameter.

	\end{abstract}
	
	\maketitle
	
	\section{INTRODUCTION}

{In hybrid neutron stars (HNSs), the remnants of massive stars, the nucleons get decomposed into quarks and therefore form deconfined strange
quark matter. The strange quark matter in these stars is surrounded by hadronic matter. Quantum chromodynamics (QCD) explains such phase transition from baryonic to deconfined quark matter at ultra-high densities \cite{Alford}.
Instead of this phase transition, HNSs can experience sequential phase transitions from hadronic matter to low- and then to high-density quark matter phases \cite{arXiv:2301.10940}.
Depending on the energy density jump at the phase transition interface, the liquid quark core may reach an elastic hadronic phase or a liquid hadronic phase which this determines the ellipticity in HNSs \cite{arXiv:2210.14048}.
Hadron-quark conversion speed at the sharp interface can be slow which this can lead to a new class of dynamically stable hybrid stars \cite{arXiv:2106.10380}.
Moreover, the deconfinement phase transition in HNSs may be weakened by the axion fields
resulting in the stabilization of the massive HNSs against gravitational collapse \cite{arXiv:2206.01631}.
Astrophysical observations of compact stars are in agreement with the results for the HNSs within the MIT bag model \cite{arXiv:2107.08971}, QCD motivated chiral approach \cite{arXiv:1411.2856},
vector bag model \cite{arXiv:2303.06387}, self-consistent Nambu-Jona-Lasinio (NJL) model \cite{arXiv:1403.7492,arXiv:1511.08561,arXiv:1703.01431,arXiv:1804.10785,arXiv:2004.07909,arXiv:2104.01519}, and Field Correlator Method \cite{arXiv:1909.08661}.
}
{Different models describing the quark matter have been considered to explain HNSs \cite{arXiv:1503.02795,arXiv:1610.06435,arXiv:2105.06239,arXiv:2112.09595,arXiv:2303.04653,arXiv:2305.01246}.
Applying the Dyson-Schwinger quark model, the global reduction of the effective interaction as a result of the quark-gluon vertex and gluon propagator
alters the phase transition, the equation of state (EoS), the mass, and radius of HNS \cite{arXiv:1503.02795}.
NJL quark model predicts the effects of the vector-isoscalar terms on the mass of massive HNSs and the influence of the
vector-isovector terms on the size of the quark cores in these stars \cite{arXiv:1610.06435}.
The SU(3) NJL model which incorporates four and eight vector interactions in the quark matter, can forecast the strangeness content of the quark core in HNSs \cite{arXiv:2105.06239}.
In the NJL model, the microscopic vector interactions also act on the quark matter significantly, leading to the stiffening of the HNS EoS and growing the star mass, while
the nucleon-quark transition gently alters the star radius \cite{arXiv:2112.09595,arXiv:2305.01662}.
In the bag model, the bag pressure has important effects on the special points in the mass radius relation of HNSs \cite{arXiv:2303.04653}.
In the quasiparticle model, the coupling constant results in the stiffening of the HNS EoS which this enhances the HNS maximum mass and reduces
the quark core mass and radius \cite{arXiv:2305.01246}.
}
{Various aspects of the HNS EoS as a factor that importantly affects the structural properties of HNSs, have been probed in the literature.
The Walceka model and the self-consistent NJL model lead to the HNS EoSs that result in the self-bound stars \cite{arXiv:2104.01519}.
Symmetry energy, as an important part of the EoS in strongly interacting matter and quark matter, is constrained via the observed maximum masses of HNSs \cite{arXiv:2202.11463}.
Parameter space of the HNS EoSs can be constrained for the nucleonic and quark matter parts employing the nuclear calculations, multi-messenger astrophysical data, and perturbative QCD \cite{arXiv:2210.09077}.
The dark matter in HNSs which interacts with hadronic and quark matter via the exchange of the Higgs boson,
alters the discontinuity on the energy density of HNSs and changes the star mass radius relation, the minimum mass, and the quark core radius of these stars \cite{arXiv:2212.12615,fuzzy}.
The EoS of HNSs which contain the quark matter core and nuclear outer phase satisfies about 500 relativistic mean-field theories constrained by chiral effective field theory and properties of isospin-symmetric nuclear matter \cite{arXiv:2302.02989}.
Astrophysical and gravitational wave observations approve the softening of matter in the intermediate density range and its stiffening at higher densities,
unlike the pure hadronic EoS \cite{arXiv:2303.06387}.
Constant-Sound-Speed parametrization for the quark matter EoS predicts the lower transition energy density, the smaller energy density discontinuity, and the
higher sound speed of quark matter in HNSs \cite{arXiv:2308.06993}.
}
{Scalar tensor gravity as one of the most favorable modified theories of gravity has been progressed to explain the recent observational data against
the general relativity (GR). In this theory, the tachyonic instability as a result of the scalar field and curvature
nonminimal coupling results in the spontaneous scalarization in relativistic stars.
This spontaneous scalarization in stars may be caused by the coupling to curvature invariants as well as by the scalar-gauge coupling in Einstein gravity \cite{arXiv:2105.14661}.
The neutron star mass and radius measurements can constrain the spontaneous scalarization parameters through the Bayesian analysis \cite{arXiv:2204.02138}.
In addition to the mass and radius of compact stars, the star scalar field also alters the overall shape of pulse profile in neutron stars leading to the opportunity
for study the curvature of spacetime \cite{arXiv:1808.04391}.
Another property of neutron stars altered by the scalar tensor gravity is the maximum compactness of scalarized stars which is lower than the non-scalarized ones \cite{arXiv:1806.00568}.
Moreover, the magnetic and the scalarization aspects of neutron stars depend on the interchange between magnetic and scalar fields of stars \cite{arXiv:2005.12758}.
On the other, the spontaneous scalarization of neutron stars changes the star magnetic deformation and quadrupolar gravitational wave emissions \cite{arXiv:2010.14833}.
}
{Scalarized compact stars have been also investigated in modified scalar tensor theories of gravity such as massive scalar tensor
gravity \cite{arXiv:1805.07818,arXiv:1812.00347,arXiv:2109.13453}, tensor multi scalar theories of gravity \cite{arXiv:1909.00473,arXiv:1911.06908,arXiv:2004.03956},
scalar Gauss-Bonnet gravity \cite{arXiv:2103.11999,arXiv:2111.03644,arXiv:2212.07653}, and degenerate higher order scalar tensor theories \cite{arXiv:2207.13624}.
The universal relations for the moment of inertia - mass of scalarized neutron stars in scalar tensor gravity with self-interacting massive scalar field express
important deviations from the general relativistic ones \cite{arXiv:1805.07818,arXiv:1812.00347}.
The massive scalar tensor theories can be constrained using the X-ray pulse profiles, the tidal deformability, and the universal relation of scalarized neutron stars \cite{arXiv:2109.13453}.
In massive tensor multi scalar theories of gravity, there exists mixed configurations of tensor multi scalar solitons
and neutron stars with larger central values of the scalar field in comparison with the pure neutron stars \cite{arXiv:1909.00473}.
Tensor multi scalar theories of gravity predict the topological neutron stars which are
specified by a topological charge \cite{arXiv:1911.06908}.
Besides, non-topological spontaneously scalarized neutron stars have been explored in the tensor multi scalar theories of gravity \cite{arXiv:2004.03956}.
Scalar Gauss-Bonnet gravity permits the spontaneous scalarization of black holes and neutron stars via the spherically
symmetric nonlinear stellar core collapse \cite{arXiv:2103.11999}.
In scalar Gauss-Bonnet gravity, the scalar field couplings to Ricci scalar and Gauss-Bonnet invariant affect the domain
of existence and the amount of scalar charge in scalarized neutron stars \cite{arXiv:2111.03644}.
Considering the black holes in the Teleparallel gravity with a coupling between scalar field and the Gauss-Bonnet invariant,
the scalarization is produced by the torsion in these objects \cite{arXiv:2212.07653}.
Furthermore, scalarized neutron stars in degenerate higher order scalar tensor theories have masses and radii larger than the ones in GR \cite{arXiv:2207.13624}.
In this work, we investigate the properties of scalarized HNSs applying different quark matter models and HNS EoSs in scalar tensor gravity. In section \ref{s2}, the EoSs which describe the HNSs are presented.
Section \ref{s3} is related to scalar tensor theory. The structure of scalarized HNSs is explained in section \ref{s4}. Section \ref{s5} is devoted to conclusion.}

\section{Hybrid Neutron Star Equation of State}\label{s2}

In this paper, we utilize a model for HNS similar the one described in Ref. \cite{Pereira et al. (2021)}.
The HNS includes a quark core and a hadronic layer.
In this model, we suppose that a sharp phase transition surface without a mixed phase divides two parts and
the density can be discontinuous at the phase splitting surface \cite{Pereira et al.2020}.
For the hadronic phase, we employ the dense neutron star matter EoS in the form of a piecewise polytropic expansion constrained by the observational data related to GW170817 as well as the data of six low-mass X-ray binaries with thermonuclear
burst or the symmetry energy of the nuclear interaction \cite{Jiang et al.2019}.
In order to describe the quark core, we apply different MIT bag models including massless quark approximation \cite{Banerjee21}, massive strange quark matter considering the QCD coupling constant \cite{Farhi,Zdunik,0901.1268}, and interacting strange quark matter specified by effective bag constant and perturbative QCD corrections term \cite{Farhi}. These models for the quark phase will be described in the following.
We take the density jump at the quark-hadronic phase transition surface
as a free parameter. This jump is described by the parameter $\eta$ defined as
\begin{eqnarray}
        \eta\equiv\frac{\epsilon_q}{\epsilon_h}-1,
 \end{eqnarray}
with the density at the top of the quark phase, $\epsilon_q$, and the density at the bottom of the hadronic phase, $\epsilon_h$.
We can calculate $\epsilon_q$ or $\epsilon_h$ and the pressure at the phase transition surface using $p_q = p_h$ at the phase-splitting surface.
{The higher values of $\eta$ correspond to larger values of $\epsilon_q$ compared to $\epsilon_h$ and $\eta=0.0$ expresses the case of $\epsilon_q=\epsilon_h$.}

\begin{figure}[h]
	\subfigure{}{\includegraphics[scale=0.85]{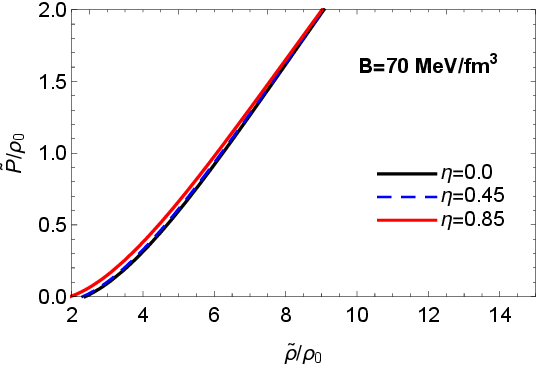}
			}
	\subfigure{}{\includegraphics[scale=0.85]{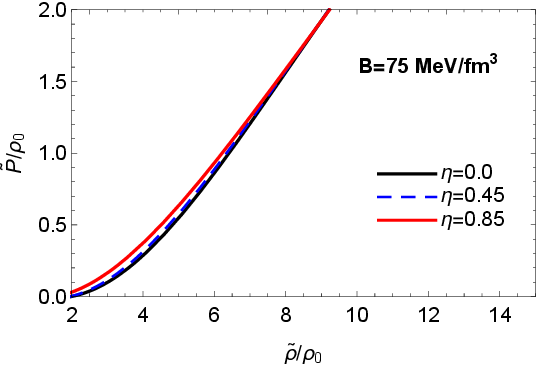}
		}	
	\caption{EoS of HNS in the massless quark approximation considering different values of the bag constant, $B$, and the density jump, $\eta$.}
	\label{p-rhot0}
\end{figure}

In the massless quark approximation, a system containing $u$, $d$, and $s$ quarks that are
noninteracting and massless is considered. Applying this approximation in MIT bag model {in which the quarks are assumed to be confined into a phenomenological
bag}, the quark pressure $P$ takes the following form
\begin{eqnarray}
       P=\frac{1}{3}(\epsilon-4B),
 \end{eqnarray}
in which $\epsilon$ denotes the energy density of the quark matter distribution and $B$ presents the bag constant that acts as the inward pressure appropriate to confine quarks inside the bag. {We note that at zero pressure, the condition $\epsilon=4B$ is satisfied.} Fig. \ref{p-rhot0} shows the EoS of HNS in the massless quark approximation with different values of the bag constant $B$ and the density jump $\eta$. We have assumed $\rho_0=1.66\times10^{14}g/cm^3$. The pressure takes higher values as the density jump increases, especially at lower densities. { Therefore, the larger density of the quark matter at the phase-splitting surface leads to the stiffer EoS of HNS.} The pressure decreases as the bag constant increases from $B=70 MeV/fm^3$ to $B=75 MeV/fm^3$.
{Consequently, higher inward pressure of quarks inside the bag gives rise to the softer EoSs.}

\begin{figure}[h]
	\subfigure{}{\includegraphics[scale=0.85]{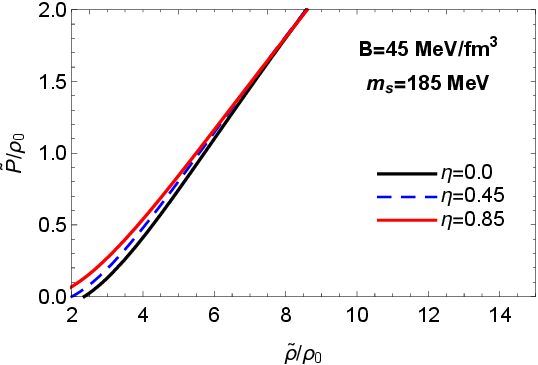}
		}	
\subfigure{}{\includegraphics[scale=0.85]{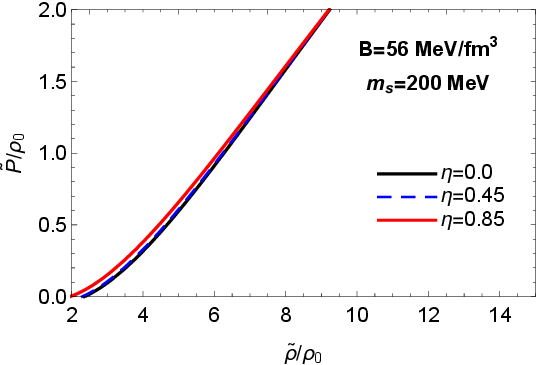}
			}
	\caption{EoS of HNS considering the massive strange quark matter with the QCD coupling constant. Different values of the bag constant, $B$, the mass of the strange quark, $m_s$, and the density jump, $\eta$, have been applied.}
	\label{p-SQM}
\end{figure}

The second MIT bag model that we employ in this work is the massive strange quark matter considering the QCD coupling constant.
In this model, the quark matter is composed of the massless $u$ and $d$ quarks, electrons $e$,
and massive $s$ quarks \cite{Farhi}. The EoS of quark matter in this system depends on the bag constant, $B$, the QCD coupling constant, $\alpha_c$, the mass of the strange quark, $m_s$, and the renormalization point, $\rho_N$. We select the values $\rho_N=313 MeV$ similar to Ref. \cite{Farhi} and $\alpha_c=0.2$ following Ref. \cite{0901.1268}. In Fig. \ref{p-SQM}, we have presented the EoS of HNS considering the massive strange quark matter with the QCD coupling constant. With lower values of the bag constant and the strange quark mass, the effect of the density jump on the EoS is more significant. Besides, the EoS is softer considering higher values of the bag constant and the strange quark mass.

\begin{figure}[h]
	\subfigure{}{\includegraphics[scale=0.85]{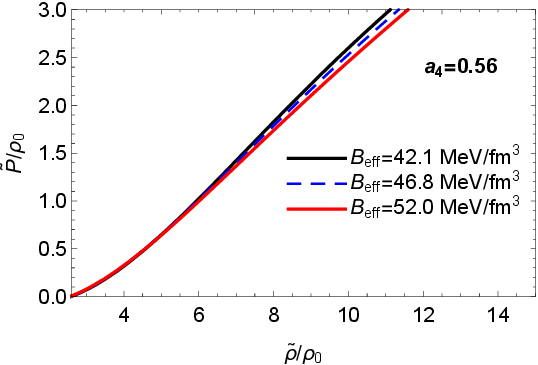}			}
	\subfigure{}{\includegraphics[scale=0.85]{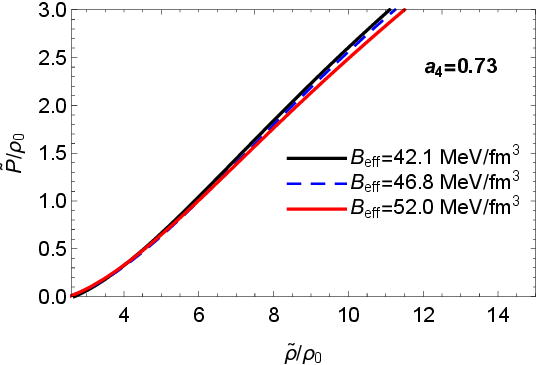}
				\subfigure{}{\includegraphics[scale=0.85]{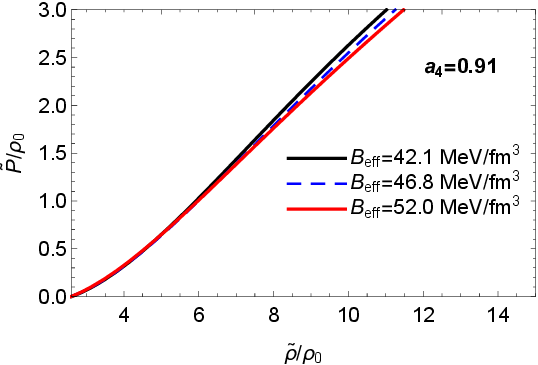}			}}	
	\caption{EoS of HNS in the third model considering different values of $B_{eff}$ and $a_4$ with fixed values of the strange quark mass $m_s = 100\ MeV$ and the density jump $\eta=0$.}
	\label{eos3}
\end{figure}

In the third model, the interacting strange quark matter characterized by the effective bag constant and the perturbative QCD corrections term is explored. For this model, a mixture of quarks, $u$, $d$, $s$, and electrons $e$, with the transformation due to weak interaction between quarks and
leptons is considered \cite{Farhi}. The grand canonical potential per unit volume is given by
\begin{eqnarray}
     \Omega=\sum_{i=u,d,s,e}\Omega_i^0+\frac{3(1-a_4)}{4\pi^2}\mu^4+B_{eff}.
 \end{eqnarray}
In the above equation, the grand canonical potential for $u$, $d$, $s$ quarks and electrons as the ideal relativistic Fermi gases is presented by
$\Omega_i^0$. Besides, the average quark chemical potential is denoted by $\mu=(\mu_u+\mu_d+\mu_s)/3$. {We present
the phenomenological representation of the non-perturbative QCD effects by the effective bag constant $B_{eff}$} and the perturbative QCD contribution from one-gluon exchange for gluon interaction by $a_4$.
The number density of particles in strange quark matter is written as
\begin{eqnarray}
           n_i=-\frac{\partial\Omega}{\partial\mu_i},
 \end{eqnarray}
in which $\mu_i(i=u,d,s,e)$ is the chemical potential of particles.
The weak interactions determine the conditions for the quark matter at the equilibrium state,
\begin{eqnarray}
          \mu_d=\mu_u+\mu_e,
 \end{eqnarray}
\begin{eqnarray}
          \mu_d=\mu_s.
 \end{eqnarray}
The charge neutrality condition is also given by
\begin{eqnarray}
         \frac{2}{3}n_u= \frac{1}{3}[n_d+n_s]+n_e.
 \end{eqnarray}
The pressure of quark matter is then as follows
 \begin{eqnarray}
        P=- \Omega,
 \end{eqnarray}
 and the energy density is given by
 \begin{eqnarray}
        \varepsilon= \Omega+\sum_{i=u,d,s,e} \mu_i n_i.
 \end{eqnarray}
Fig. \ref{eos3} shows the EoS of HNS in the third model with different values of $B_{eff}$ and $a_4$.
We have fixed the strange quark mass $m_s = 100\ MeV$ and the density jump $\eta=0$. The pressure decreases by increasing the effective bag constant.
{Thus, this non-perturbative effect of the strong interactions results in the softening of the EoS.}
The influence of the effective bag constant is more significant at higher densities.
{Our calculations verify that the perturbative QCD correction, $a_4$, does not significantly affect the HNS EoS.} In this paper, we investigate the scalarized HNSs which described by the above EoSs.

\section{Scalar Tensor gravity}\label{s3}
Scalar tensor theories in Einstein frame can be defined by the following action
\begin{equation}
	S[g_{\mu \nu},\phi,\Psi_m] = \frac{1}{16 \pi} \int d^4x \sqrt{-g}(R-2 \nabla_\mu\phi \nabla^\mu \phi)+S_m[\Psi_m,a(\phi)^2 g_{\mu \nu}].
\end{equation}
In the above equation, $g=det(g_{\mu \nu})$, $R$ is Ricci scalar, $\phi$ shows the scalar field, $\Psi_m$ denotes the matter field, and $a(\phi)$ is the coupling function satisfying the relation $\tilde{g} _{\mu \nu}=a(\phi)g_{\mu \nu}$ between $g_{\mu \nu}$ in Einstein frame and $\tilde{g_{\mu \nu}}$ in Jordan frame.
Here, we assume the form of the coupling function as $a(\phi) = e^{\frac{1}{2}\beta (\phi-\phi_0) ^2}$ with the coupling constant $\beta$ and  $\phi_0=0$.
To describe a spherical symmetric static HNS in the scalar tensor theory in Einstein frame, we apply the following form of the spacetime line element
\begin{equation}
	ds^2 = - N(r)^2 dt^2 + A(r)^2 dr^2 + r^2
	(d\theta^2 + \sin^2\theta d\phi^2),
\end{equation}
with the metric functions $N(r)$ and $A(r)= [1-2 m(r)/r ]^{-1/2}$ and the mass profile $m(r)$. Therefore, the field equations result in the
following differential equations [28],
\begin{align}
	&\frac{d m}{dr} = 4\pi r^2 a^4 \tilde{\epsilon} + \frac{r}{2} (r-2m) \Big(\frac{d\phi}{dr}\Big)^2 \label{eq:dm},\\
	&\frac{d \ln N}{dr} = \frac{4\pi r^2 a^4 \tilde{p}}{r - 2m} +\frac{r}{2} \Big(\frac{d\phi}{dr}\Big)^2 + \frac{m}{r(r-2m)} \label{eq:dn}, \\
	&\frac{d^2\phi}{dr^2} = \frac{4\pi r a^4}{r-2m} \! \left[ \alpha (\tilde{\epsilon} - 3\tilde{p}) + r (\tilde{\epsilon} - \tilde{p}) \frac{d\phi}{dr} \right ]\! -\frac{2(r-m)}{r(r-2m)} \frac{d\phi}{dr} \label{eq:dphi}, \\
	&\frac{d\tilde{p}}{dr} = -(\tilde{\epsilon} + \tilde{p}) \left[  \frac{4\pi r^2 a^4 \tilde{p}}{r-2m} \! + \! \frac{r}{2} \Big(\frac{d\phi}{dr}\Big)^2 \!\! + \! \frac{m}{r(r-2m)} \! + \! \alpha \frac{d\phi}{dr} \right], \label{eq:dp}\\
	&\frac{dm_b}{dr}=\frac{4\pi r^2a^3\tilde{\rho}}{\sqrt{1-\frac{2m}{r}}}\label{eq:dm_b}.
\end{align}
Here, $\tilde{\epsilon}$ is the energy density, $\tilde{p}$ shows the pressure, and $m_b$ denotes the baryonic mass.
The boundary conditions which used for solving these equations are
\begin{align}
	&m(0) =m_b(0)= 0, \quad \lim_{r\to\infty}N(r) = 1,\quad \phi(0)=\phi_c, \quad \lim_{r\to\infty}\phi(r) = 0, \nonumber \\
	&\frac{d\phi}{dr}(0) = 0, \qquad \tilde{p}(0) = p_c, \qquad \tilde{p}(R_s) = 0. \label{eq:bc}
\end{align}
Here, the star radius is presented by $R_s$ and $c$ shows the center of star.
Considering an appropriate boundary conditions at $r = 0$ as well as a guess
$\phi(0)=\phi_c$ at the center, the iteration on $\phi_c$ is done getting the condition \cite{26,28},
\begin{equation} \label{eq:constraint on central of scalar field}
	\phi_s  + \frac{2 \psi_s}{\sqrt{\dot{\nu}_s^2+4\psi_s^2}} \textrm{arctanh} \left[ \frac{\sqrt{\dot{\nu}_s^2 +4\psi_s^2}}{\dot{\nu}_s +2/R_s} \right] = 0,
\end{equation}
in which s indicates the quantities on the surface of the star as well as $\psi_s = (d\phi/dr)_s$ and $\dot{\nu}_s = 2(d\ln N/dr)|_s = R_s \psi_s^2 + 2 m_s/[R_s(R_s-2m_s)]$. The ADM mass, $M_{ADM}$, and the scalar charge, $\omega$, are also calculated as follows [26,28],
\begin{align}
	M_{ADM} &= \frac{R_s^2 \dot{\nu}_s}{2} \left( 1-\frac{2m_s}{R_s} \right)^\frac{1}{2}
	\exp \left[ \frac{-\dot{\nu}_s}{\sqrt{\dot{\nu}_s^2+4\psi_s^2}} \textrm{arctanh} \left( \frac{\sqrt{\dot{\nu}_s^2+4\psi_s^2}}{\dot{\nu}_s +2/R_s} \right) \right], \\
	\omega & = - 2 M_{ADM} \psi_s/\dot{\nu}_s.
\end{align}
Here, we report the values of the HNS ADM mass. In the following, we study the HNS in the scalar tensor gravity.

\section{RESULTS and DISCUSSION}\label{s4}
\subsection{Mass versus the density of scalarized hybrid neutron star}\label{subsec:M-rho}

\begin{figure}[h]
	\subfigure{}{\includegraphics[scale=0.85]{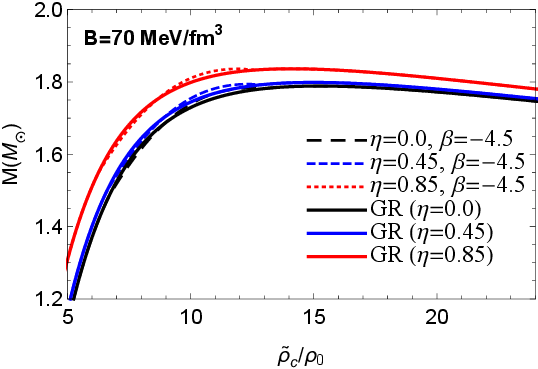}
		}
	\subfigure{}{\includegraphics[scale=0.85]{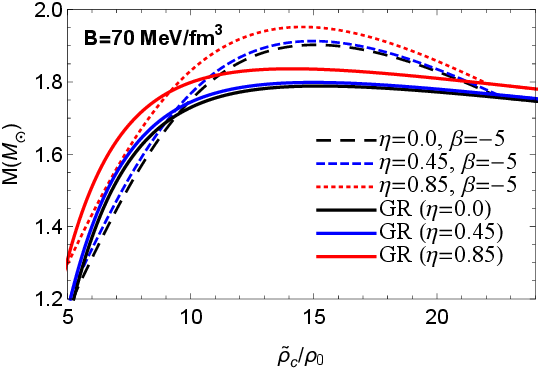}
		
		\subfigure{}{\includegraphics[scale=0.85]{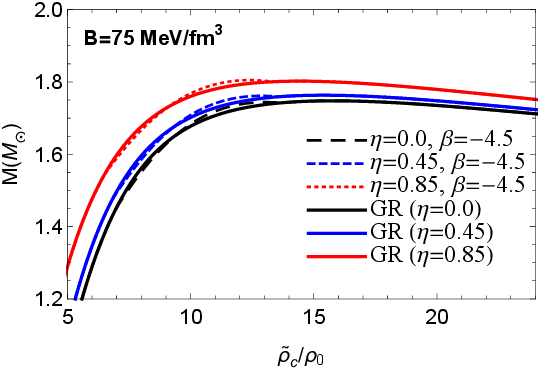}
			}
		\subfigure{}{\includegraphics[scale=0.85]{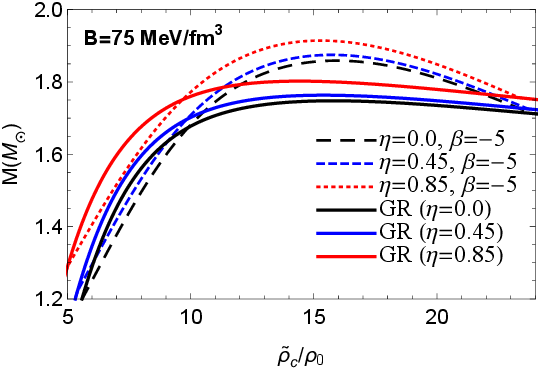}
			}
		}	
	\caption{HNS mass in the first model for the HNS EoS applying the scalar tensor gravity and GR.
Different values of the bag constant, $B$, the density jump, $\eta$, and the coupling constant, $\beta$, have been considered.}
	\label{M1}
\end{figure}

\begin{figure}[h]
		\subfigure{}{\includegraphics[scale=0.85]{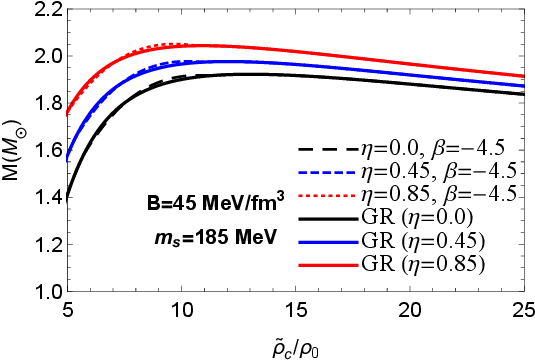}
		}
	\subfigure{}{\includegraphics[scale=0.85]{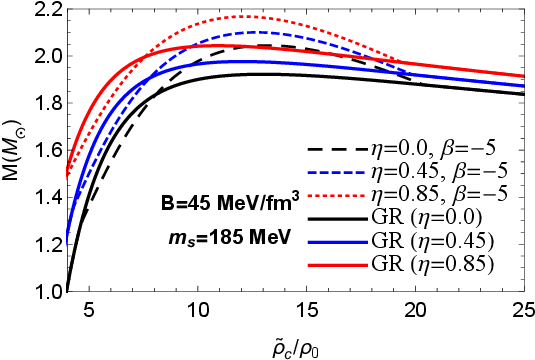}
		}	
\subfigure{}{\includegraphics[scale=0.85]{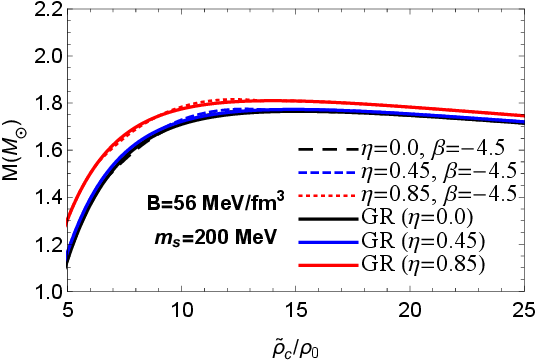}
		}
	\subfigure{}{\includegraphics[scale=0.85]{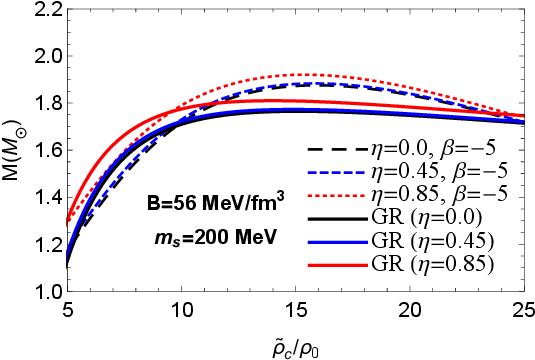}
		}
	\caption{Same as Fig. \ref{M1} but for the second model of the HNS EoS. We have applied different values of the bag constant, $B$, the mass of the strange quark, $m_s$, and the density jump, $\eta$.}
	\label{M2}
\end{figure}

\begin{figure}[h]
	\subfigure{}{\includegraphics[scale=0.85]{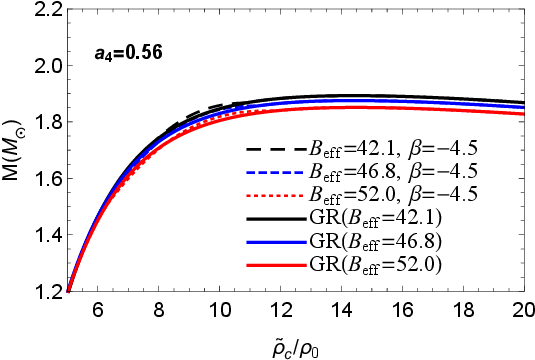}
		}
	\subfigure{}{\includegraphics[scale=0.85]{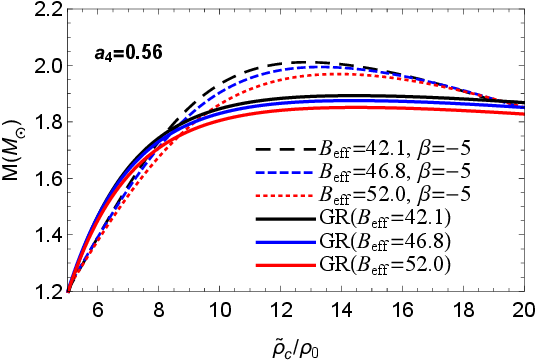}
		}
	\subfigure{}{\includegraphics[scale=0.85]{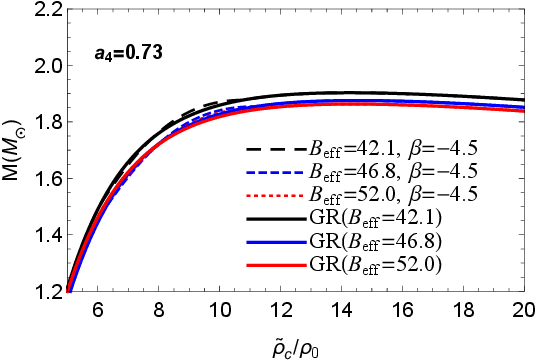}
		}
	\subfigure{}{\includegraphics[scale=0.85]{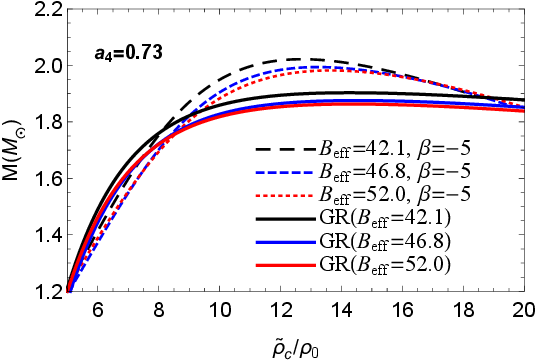}
		}
	\subfigure{}{\includegraphics[scale=0.85]{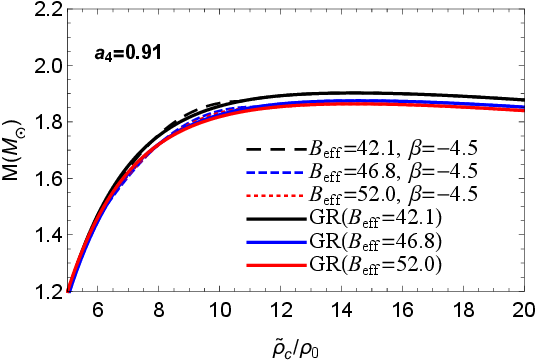}
		}
	\subfigure{}{\includegraphics[scale=0.85]{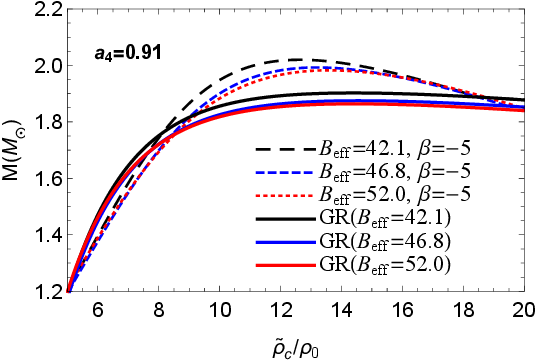}
		}
	\caption{Same as Fig. \ref{M1} but for the third model of the HNS EoS.
Different values of $B_{eff}$ (in units of $\frac{MeV}{fm^3}$) and $a_4$ with fixed values of the strange quark mass, $m_s = 100\ MeV$, and the density jump, $\eta=0$, have been considered.}
	\label{M3}
\end{figure}

We have presented the mass of HNS applying different EoS models in Figs. \ref{M1}-\ref{M3}.
Figs. \ref{M1} and \ref{M2} confirm that with the higher values of the density jump, the star mass grows. This effect becomes more important by increasing $\eta=0.45$ to $\eta=0.85$. {Hence, the quark cores with larger densities at the phase-splitting surface increase the star mass.}
{We find from Fig. \ref{M2} that the increase in $B$ and $m_s$ which causes the softening of the EoS, reduces the maximum mass of scalarized stars.}
Fig. \ref{M3} verifies that the star mass reduces as $B_{eff}$ increases {due to the softening of the EoS by $B_{eff}$}. In all cases, the deviation of scalarized star mass from the GR one is more clear with smaller values of $\beta$ {as expected from the scalar tensor gravity with negative coupling constants.}
Moreover, the range of star mass at which the HNS is scalarized becomes larger by decreasing the coupling constant. Applying three models for the HNS EoS, the most massive stars are the scalarized ones {in agreement with the results of Ref. \cite{28}}. With lower values of $\beta$, the maximum star mass grows. The central densities corresponding to most massive scalarized stars decrease as the density jump increases. {This means that in the case of the quark cores with the higher densities at the phase transition surface, the HNS can be easily more massive.} {Our results show that by considering the higher values of the bag constant, i.e. softer EoSs, the scalarization of the HNSs continues to larger central densities (see Figs. \ref{M1} and \ref{M2}).}
It is obvious from Fig. \ref{M3} that the most massive stars in the third model are the scalarized ones with lower values of $B_{eff}$ {(stiffer EoSs)} and $\beta$. Besides, the central density related to the maximum mass of scalarized stars in the third model increases by $B_{eff}$.


In Tables \ref{table1}-\ref{table3}, the maximum mass of HNS considering different models for the HNS EoS employing the scalar tensor gravity and GR is presented. These tables confirm that the maximum mass of scalarized stars with $\beta=-4.5$ is almost equal to the GR one. {These stars are less affected by the scalar tensor gravity.} Besides, for both scalarized stars and the stars in GR, the maximum mass decreases by increasing the coupling constant. Table \ref{table2} shows that in the second model, the maximum mass decreases by increasing the bag constant and the mass of the strange quark, { because of the softening of the EoS.} In addition, it is obvious from Table \ref{table3} that the maximum mass of HNS is not significantly affected by the model parameter $a_4$.

\begin{table}[h!]
	\begin{center}
		\begin{tabular}{|c@{\hspace{3mm}}|c@{\hspace{3mm}}c@{\hspace{3mm}}c@{\hspace{3mm}}|c@{\hspace{3mm}}c@{\hspace{3mm}}c@{\hspace{3mm}}|}
			\hline
						 & &$B=70 MeV/fm^3$ & & &$B=75 MeV/fm^3$& \\	
			\hline
			$\beta$ & $\eta=0$&$\eta=0.45$&$\eta=0.85$&$\eta=0$&$\eta=0.45$&$\eta=0.85$\\	
			\hline
			{$-4.5$}&1.79 &1.80 &1.84 &1.75 &1.76 &1.80   \\
		{$-5$}&1.90 &1.91 &1.95 &1.86 &1.87 &1.91   \\
		{$GR$}&1.79 &1.80 &1.84 &1.75 &1.76 &1.80   \\
			\hline\hline

		\end{tabular}
		
	\end{center}
	\caption{Maximum mass (in solar mass unit) of HNS in the first model for the HNS EoS with different values of the bag constant, $B$, the density
jump, $\eta$, and the coupling constant, $\beta$.}
	\label{table1}
\end{table}

\begin{table}[h!]
	\begin{center}
		 \begin{tabular}{|c@{\hspace{2mm}}|c@{\hspace{2mm}}c@{\hspace{2mm}}c@{\hspace{3mm}}|c@{\hspace{3mm}}c@{\hspace{3mm}}c@{\hspace{3mm}}|}
			
			\hline
			& &$B=45 MeV/fm^3$&&&$B=56 MeV/fm^3$&\\
& &$m_s=185 MeV$&&&$m_s=200 MeV$ &\\
\hline
			$\beta$& $\eta=0$&$\eta=0.45$&$\eta=0.85$&$\eta=0$&$\eta=0.45$&$\eta=0.85$\\	
			\hline
			{$-4.5$}&1.92 &1.98 &2.05&1.76 &1.77 &1.82    \\
			{$-5$} &2.04 &2.10 &2.17 &1.88 &1.88 &1.92 \\
			{$GR$} &1.92 &1.98 &2.04&1.76 &1.77 &1.81   \\
			\hline\hline
		\end{tabular}
		
	\end{center}
	\caption{Same as Table \ref{table1} but for the second model of the HNS EoS.
	}
	\label{table2}
\end{table}

\begin{table}[h!]
	\begin{center}
		 \begin{tabular}{|c@{\hspace{2mm}}|c@{\hspace{2mm}}|c@{\hspace{2mm}}c@{\hspace{2mm}}c@{\hspace{2mm}}|c@{\hspace{2mm}}c@{\hspace{2mm}}c@{\hspace{2mm}}|c@{\hspace{2mm}}c@{\hspace{2mm}}c@{\hspace{2mm}}|}
			
			\hline
						&& &$a_4=0.56$&&&$a_4=0.73$&&&$a_4=0.91$&\\	
\hline
&$B_{eff} (\frac{MeV}{fm^3})$& $42.1$&$46.8$&$52.0$&$42.1$&$46.8$&$52.0$&$42.1$&$46.8$&$52.0$\\
$\beta$& & &&&&&&&&\\
			\hline
			{$-4.5$}&& 1.89 &1.88 &1.85&1.90 &1.88 &1.86&1.90 &1.87 &1.86   \\
			{$-5$}&& 2.01 &1.99 &1.97&2.02 &1.99 &1.98&2.02 &1.99 &1.98   \\
			{$GR$}&& 1.89 &1.88 &1.85 &1.90 &1.88 &1.86 &1.90 &1.87 &1.86  \\
			\hline\hline
		\end{tabular}
		
	\end{center}
	\caption{Same as Table \ref{table1} but for the third model of the HNS EoS.	}
	\label{table3}
\end{table}

\subsection{Mass radius relation of scalarized hybrid neutron star}

\begin{figure}[h]
	\subfigure{}{\includegraphics[scale=0.85]{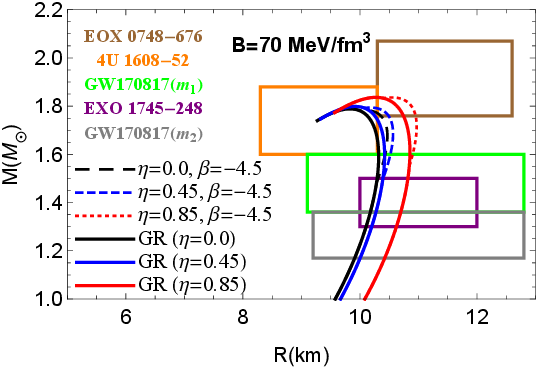}
		}
	\subfigure{}{\includegraphics[scale=0.85]{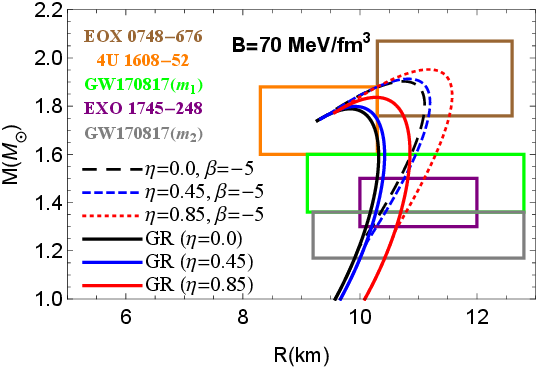}
		
		\subfigure{}{\includegraphics[scale=0.85]{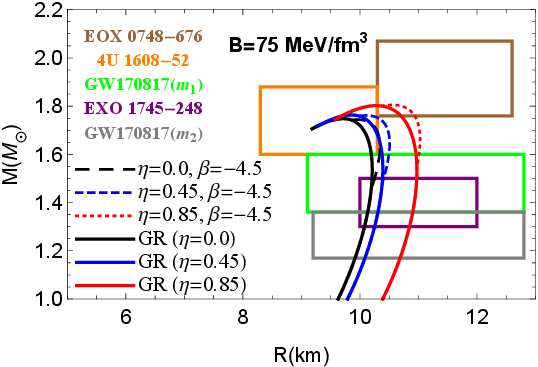}
			}
		\subfigure{}{\includegraphics[scale=0.85]{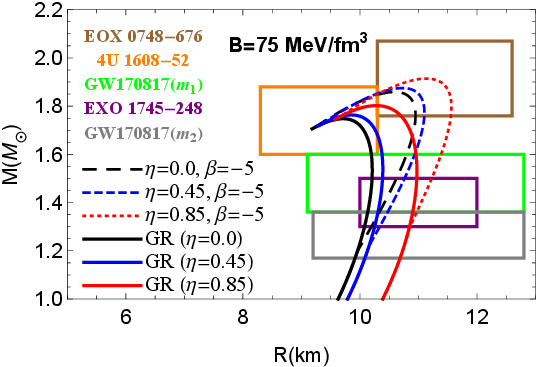}
			}
		}	
	\caption{Mass radius relation in the first model for the HNS EoS applying the scalar tensor gravity and GR.
Different values of the bag constant, $B$, the density jump, $\eta$, and the coupling constant, $\beta$, have been considered. We have also shown the observational constraints on the radii and masses of compact stars, EOX 0748-676 \cite{Cheng2017}, 4U 1608-52 \cite{Guver2010}, GW170817 \cite{Abbott7,Abbott8}, and EXO 1745-248 \cite{Ozel2009}.
}
	\label{Mr1}
\end{figure}

\begin{figure}[h]
		\subfigure{}{\includegraphics[scale=0.85]{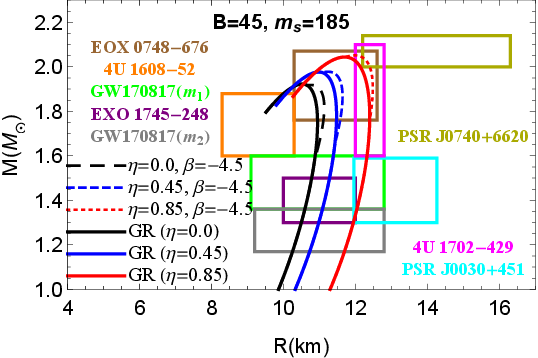}
		}
	\subfigure{}{\includegraphics[scale=0.85]{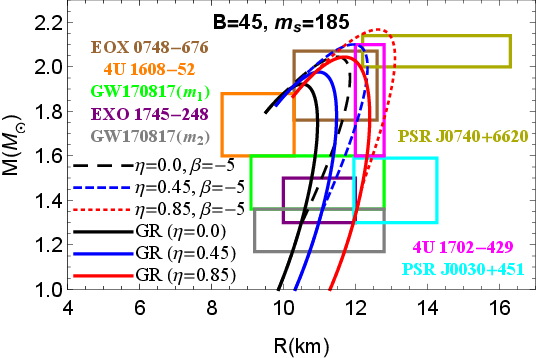}
		}	
\subfigure{}{\includegraphics[scale=0.85]{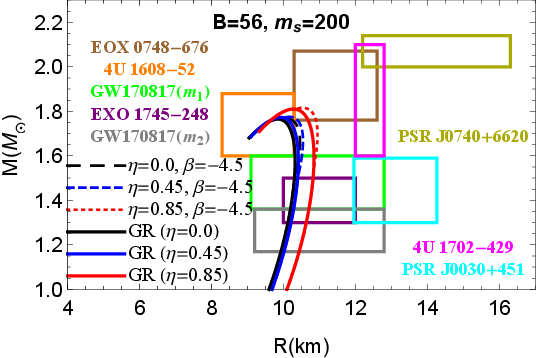}
		}
	\subfigure{}{\includegraphics[scale=0.85]{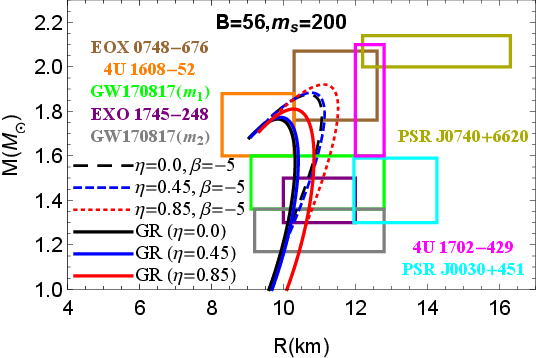}
		}
	\caption{Same as Fig. \ref{Mr1} but for the second model of the HNS EoS. We have applied different values of the bag constant, $B$, the mass of the strange quark, $m_s$, and the density jump, $\eta$. $B$ and $m_s$ are in units of $\frac{MeV}{fm^3}$ and $MeV$, respectively. In addition to the constraints given in Fig. \ref{Mr1}, the observational constraints related to PSR J0740+6620 \cite{Fonseca,Miller}, 4U 1702-429 \cite{Nattila}, and PSR J0030+451 \cite{Miller2019} have also been presented.}
	\label{Mr2}
\end{figure}

\begin{figure}[h]
	\subfigure{}{\includegraphics[scale=0.85]{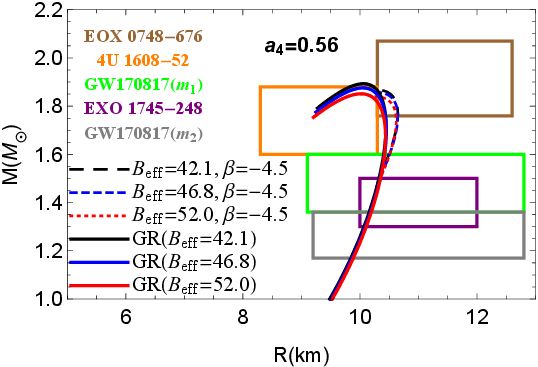}
		}
	\subfigure{}{\includegraphics[scale=0.85]{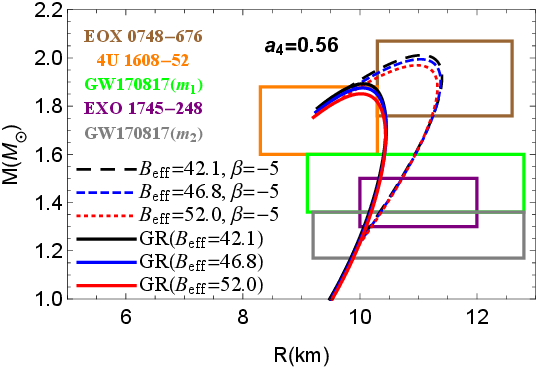}
		}
	\subfigure{}{\includegraphics[scale=0.85]{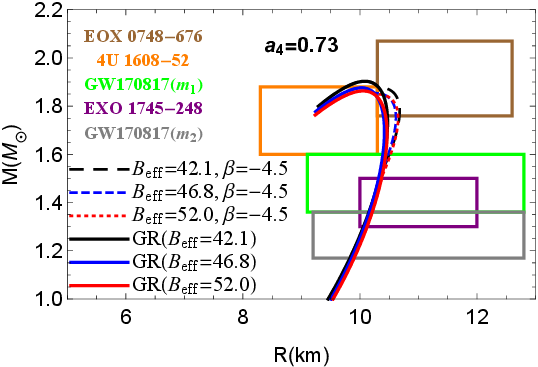}
		}
	\subfigure{}{\includegraphics[scale=0.85]{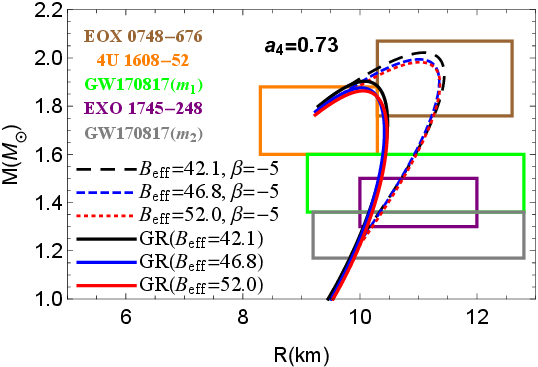}
		}
	\subfigure{}{\includegraphics[scale=0.85]{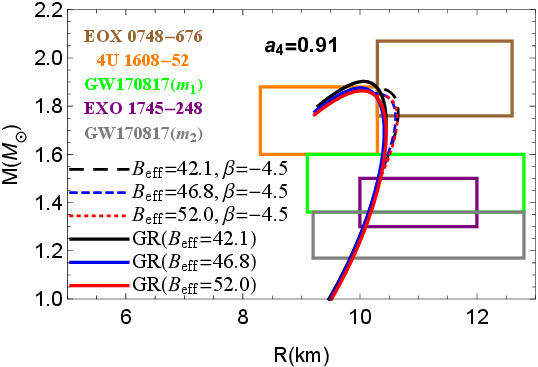}
		}
	\subfigure{}{\includegraphics[scale=0.85]{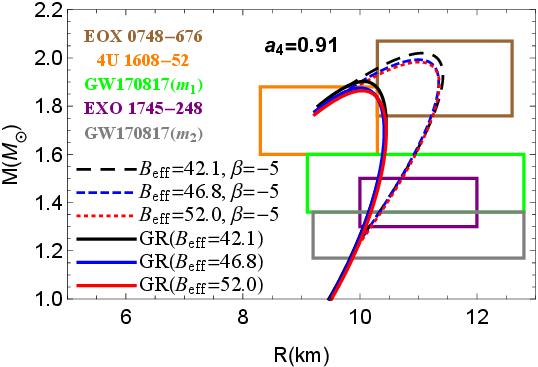}
		}
	\caption{Same as Fig. \ref{Mr1} but for the third model of the HNS EoS.
Different values of $B_{eff}$ (in units of $\frac{MeV}{fm^3}$) and $a_4$ with fixed values of the strange quark mass, $m_s = 100\ MeV$, and the density jump, $\eta=0$, have been considered.}
	\label{Mr3}
\end{figure}

{The mass radius relation for HNS in different EoS models has been given in Figs. \ref{Mr1}-\ref{Mr3}.
We can find from Figs. \ref{Mr1} and \ref{Mr2} that the mass and radius of stars grow by the density jump. This is due to the fact that
the quark core with larger density at the phase-splitting surface (stiffer EoS of HNS) increases the HNS mass and radius. This is while the softening of the EoS because of the growth in $B_{eff}$ reduces the mass and radius of stars (see Fig. \ref{Mr3}). This influence of $B_{eff}$ on the mass radius relation is more clear in massive stars. Figs. \ref{Mr1}-\ref{Mr3} confirm that in all EoS models, the scalarized stars are larger in size compared to GR ones. The deviation of scalar tensor gravity from GR in the mass radius relation is more obvious in massive stars. With lower values of $\beta$ at which this deviation is more important and more extended, the behavior of the mass radius relation is similar to the one related to the self-bound stars in agreement with the results of Ref. \cite{arXiv:2104.01519}. Fig. \ref{Mr2} explains that in the second model of the HNS EoS,
the star mass and radius decrease by increasing the bag constant and the strange quark mass due to the softening of the EoS. In addition, the impacts of the density jump on the mass radius relation are more notable when the bag constant and the strange quark mass are smaller, i.e. stiffer EoS. Fig. \ref{Mr3} shows that because of the small effects of $a_4$ on the HNS EoS, the variation of $a_4$ does not significantly influence the mass radius relation. In Figs. \ref{Mr1}-\ref{Mr3}, we have also presented the observational constraints on the neutron star radius and mass. Our results in the first model of the HNS EoS
satisfy the constraints from EOX 0748-676, 4U 1608-52, GW170817, and EXO 1745-248 (see Fig. \ref{Mr1}). The stars related to EOX 0748-676, GW170817, and EXO 1745-248 can be scalarized HNSs of the first model. Fig. \ref{Mr2} also presents the observational constraints from PSR J0740+6620, 4U 1702-429, and PSR J0030+451. The results of the HNS mass and radius in the second model confirm the observational constraints. The objects related to EOX 0748-676, 4U 1702-429, GW170817, PSR J0740+6620, and PSR J0030+451 may be scalarized HNSs in the second model. Moreover, Fig. \ref{Mr3} affirms that the mass and radius of scalarized HNSs in the third model are consistent with the constraints from EOX 0748-676, GW170817, and EXO 1745-248.
}


\subsection{Central scalar field in hybrid neutron star}

\begin{figure}[h]
	\subfigure{}{\includegraphics[scale=0.85]{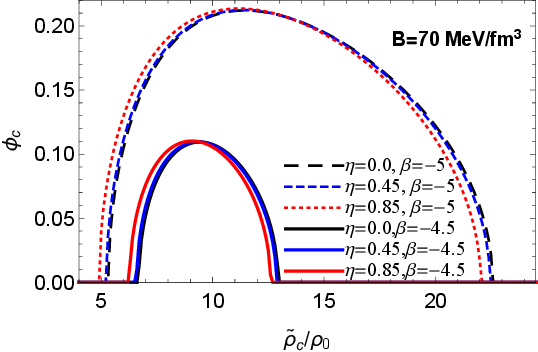}
		\label{phic-rho-11}}
	\subfigure{}{\includegraphics[scale=0.85]{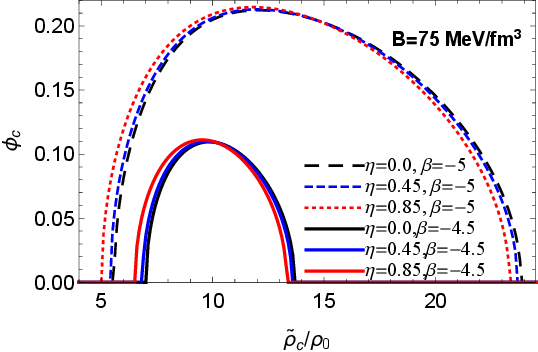}
		\label{phic-rho-22}}	
	\caption{Central scalar field, $\phi_c$, versus the central density, $\rho_c$, applying first model of the HNS EoS with different values of the bag constant, $B$, the density jump, $\eta$, and the coupling constant, $\beta$.}
	\label{phi1}
\end{figure}

\begin{figure}[h]
		\subfigure{}{\includegraphics[scale=0.85]{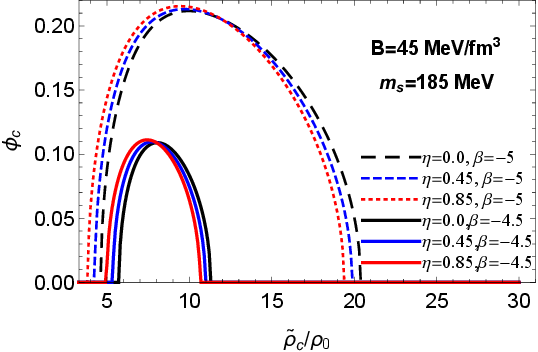}
		\label{phic-rho-2}}
\subfigure{}{\includegraphics[scale=0.85]{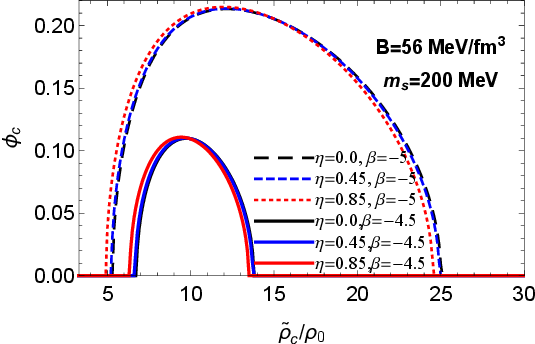}
		\label{phic-rho-1}}	
	\caption{ Same as Fig. \ref{phi1} but for the second model of the HNS EoS considering different values of the bag constant, $B$, the mass of the strange quark, $m_s$, and the density jump, $\eta$.}
	\label{phi2}
\end{figure}

\begin{figure}[h]
	\subfigure{}{\includegraphics[scale=0.85]{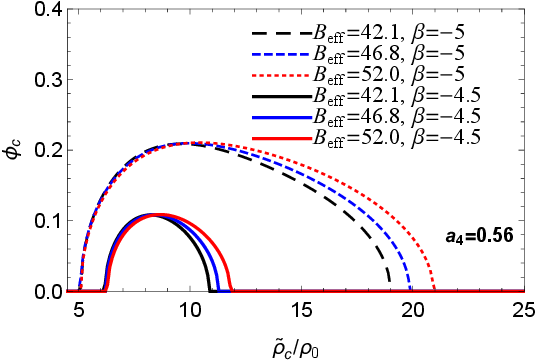}
		\label{phic-rho-1}}
	\subfigure{}{\includegraphics[scale=0.85]{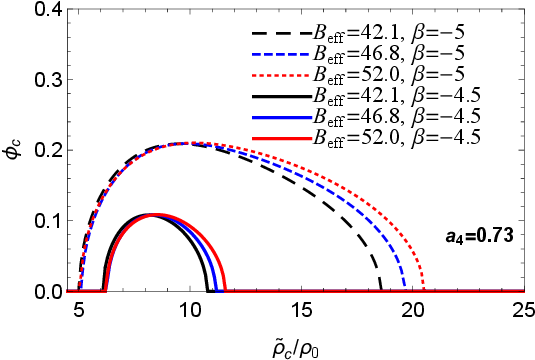}
		\label{phic-rho-2}}
	\subfigure{}{\includegraphics[scale=0.85]{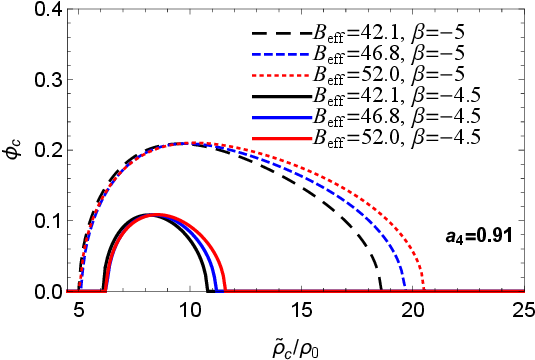}
		\label{phic-rho-1}}
	
	\caption{ Same as Fig. \ref{phi1} but for the third model of the HNS EoS applying different values of $B_{eff}$ (in units of $\frac{MeV}{fm^3}$) and $a_4$ with fixed values of the strange quark mass, $m_s = 100\ MeV$, and the density jump, $\eta=0$.}
	\label{phi3}
\end{figure}

Figs. \ref{phi1}-\ref{phi3} demonstrate the central scalar field in HNS employing different models for the HNS EoS. The coupling constant affects the spontaneous scalarization with higher values of the central scalar field considering the lower values of the coupling constant, {similar to the reported results of Ref. \cite{28}.} Furthermore, the range of the central density at which the central scalar field is nonzero becomes larger applying the smaller values of $\beta$. Figs. \ref{phi1} and \ref{phi2} verify that the central density at which $\phi_c$ becomes nonzero (first critical density) decreases as the density jump in HNS grows. {Thus, the quark core in HNS which has the larger density at the phase-splitting surface assists the appearing of the scalarization of HNS.} The central density at which the spontaneous scalarization terminates, i.e. $\phi_c=0$, (second critical density) is also smaller when the density jump is larger.
{In consequence, the larger density of the quark matter at the phase transition surface results in the disappearing of the HNS scalarization.
} Fig. \ref{phi3} shows that the effective bag constant alters the spontaneous scalarization. Our calculations indicate that the second critical density grows by increasing the effective bag constant, while the first one is not almost affected by the effective bag constant.
{This more notable influence of $B_{eff}$ at higher densities compared to low densities is based on the behavior of HNS EoSs with different effective bag constants (see Fig. \ref{eos3}).} {It is clear that the higher values of $B_{eff}$ which correspond to softer HNS EoSs keep the scalarization up to high densities.} The increase of the second critical density by $B_{eff}$ is more important when the coupling constant is smaller. In the third model of the HNS EoS, with higher values of the effective bag constant, the range of the scalarization increases. {Accordingly, the HNSs with softer EoSs due to the effective bag constant are more likely to be scalarized.} In all cases, the reduction of the coupling constant leads to the decrement in the first critical density and increment in the second one.

\begin{figure}[h]
	\subfigure{}{\includegraphics[scale=0.85]{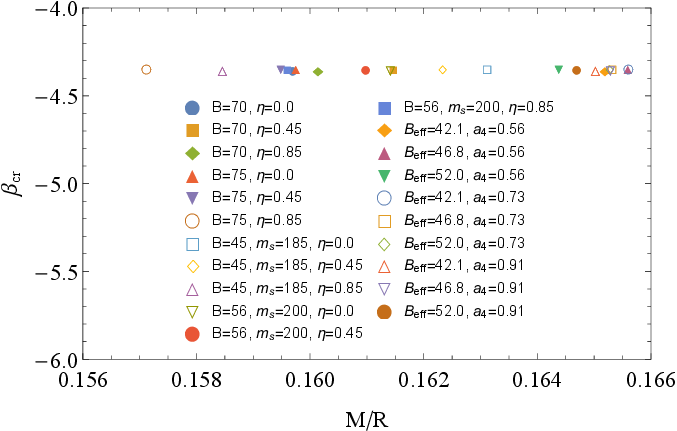}
		\label{phic-rho-1}}
	\caption{ Critical value of the coupling constant, $\beta_{cr}$, versus the star compactness, $M/R$, in different models of the HNS EoS considered in this paper. $B$ and $B_{eff}$ are in units of $\frac{MeV}{fm^3}$ and $m_s$ is in units of $MeV$.}
	\label{phi4}
\end{figure}

Tables \ref{table4}-\ref{table6} give the first critical density of the spontaneous scalarization for different models of the HNS EoS. Table \ref{table4} confirms that in the first model of the HNS EoS, the first critical density grows as the bag constant increases. This is also true for the second critical density, with higher values of the second critical density by increasing the bag constant (see Fig. \ref{phi1}). {Therefore, in the first model, the softening of the EoS by the bag constant results in the beginning and ending of the scalarization at higher densities.}
It is clear from Table \ref{table5} that in the second model of the HNS EoS, the first critical density grows as the bag constant and the mass of the strange quark increase. This enhancement also takes place for the second critical density (see Fig. \ref{phi2}). {This is also due to the softening of the EoS as a result of increase in $B$ and $m_s$.} Our results show that the effects of the bag constant and the mass of the strange quark on the spontaneous scalarization are more remarkable when the density jump in HNS is larger. {In fact, the quark core with larger density at the phase-splitting surface magnifies the effects of quark matter EoS on the spontaneous scalarization.} Table \ref{table6} approves that the influence of the effective bag constant as well as the model parameter $a_4$ on the first critical density is nearly negligible, {as expected from the weak impact of these two parameters at low densities.} Fig. \ref{phi4} presents the critical value of the coupling constant, $\beta_{cr}$, at which the spontaneous scalarization takes place. It is obvious that this critical value is nearly the same in different models of the HNS EoS, i.e. the value $\beta_{cr}\simeq -4.35$. { For this reason, we can introduce the parameter $\beta_{cr}$ as one of the model-independent parameter in scalarized HNSs.}

\begin{table}[h!]
	\begin{center}
		\begin{tabular}{
				|c@{\hspace{5mm}}|c@{\hspace{5mm}}|c@{\hspace{5mm}}|c@{\hspace{5mm}}|}
 \hline
$\beta$&$\eta$&$B=70 MeV/fm^3$&$B=75 MeV/fm^3$\\
 \cline{1-4}			
&$0$&6.7&7.1\\
 \cline{2-4}
$-4.5$&$0.45$&6.6&6.9\\
\cline{2-4}
&$0.85$&6.3&6.6\\
 \hline
&$0$&5.4&5.6\\
 \cline{2-4}
$-5$&$0.45$&5.3&5.5\\
 \cline{2-4}
&$0.85$&5&5.1\\
 \hline\hline
		\end{tabular}
	\end{center}
	\caption{First critical density (in unit of $\rho_0$) of the spontaneous scalarization in the first model of the HNS EoS.	}
	\label{table4}
\end{table}

\begin{table}[h!]
	\begin{center}
		\begin{tabular}{
				|c@{\hspace{5mm}}|c@{\hspace{5mm}}|c@{\hspace{5mm}}|c@{\hspace{5mm}}|}
 \hline
			
$\beta$&$\eta$ &$B=45 MeV/fm^3$&$B=56 MeV/fm^3$\\
& &$m_s=185 MeV$&$m_s=200 MeV$ \\
			\cline{1-4}			
			&$0$&6.4&6.8\\
			\cline{2-4}
			$-4.5$&$0.45$&5.4&6.7\\
			\cline{2-4}
			&$0.85$&5&6.4\\
			\hline
			&$0$&4.7&5.4\\
			\cline{2-4}
			$-5$&$0.45$&4.3&5.3\\
			\cline{2-4}
			&$0.85$&3.9&5\\ \hline\hline
		\end{tabular}
	\end{center}
	\caption{Same as Table \ref{table4} but for the second model of the HNS EoS.
	}
	\label{table5}
\end{table}

\begin{table}[h!]
\begin{center}
\begin{tabular}{{|c@{\hspace{5mm}}|c@{\hspace{5mm}}|c@{\hspace{5mm}}|c@{\hspace{5mm}}|c@{\hspace{5mm}}|}}
\hline
$\beta$&$B_{eff} (MeV/fm^3)$&$a_4=0.56$&$a_4=0.73$&$a_4=0.91$\\
\cline{1-5}
&$42.1$&6.2&6.2&6.2\\
\cline{2-5}
$-4.5$&$46.8$&6.3&6.3&6.3\\
\cline{2-5}
&$52.0$&6.3&6.3&6.3\\
\hline
&$42.1$&5.1&5.1&5.1\\
\cline{2-5}
$-0.5$&$46.8$&5.1&5.2&5.2\\
\cline{2-5}
&$52.0$&5.2&5.1&5.1\\
\hline\hline
	\end{tabular}
	\end{center}
	\caption{Same as Table \ref{table4} but for the third model of the HNS EoS.
	}
	\label{table6}
\end{table}

\subsection{Scalar field profile of hybrid neutron star}

\begin{figure}[h]
	\subfigure{}{\includegraphics[scale=0.85]{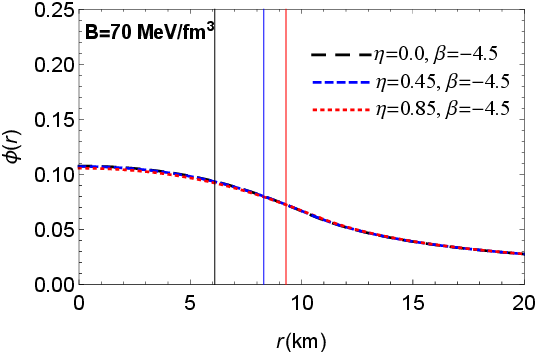}
		}
	\subfigure{}{\includegraphics[scale=0.85]{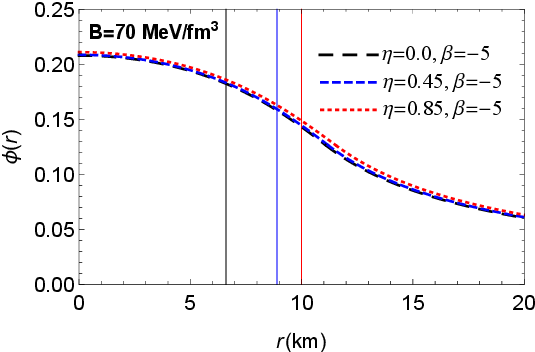}
		
		\subfigure{}{\includegraphics[scale=0.85]{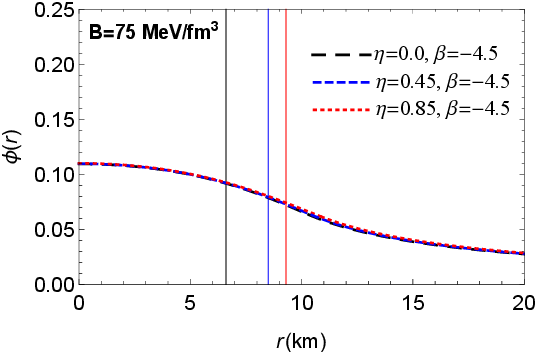}
			}
		\subfigure{}{\includegraphics[scale=0.85]{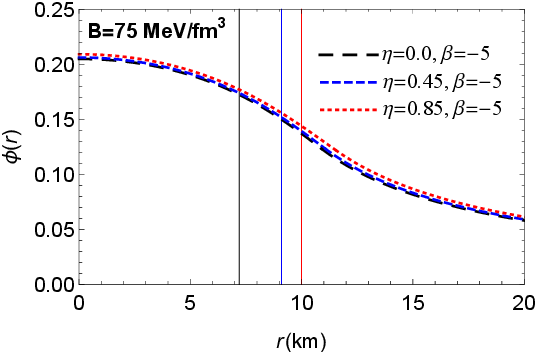}
			}
		}	
	\caption{Profile of scalar field in hybrid neutron star with the central density $\rho_c=10 \rho_0$ in the first model for the HNS EoS.
Different values of the bag constant, $B$, the density jump, $\eta$, and the coupling constant, $\beta$, have been considered. The position of the phase-splitting surface has been presented by the vertical lines, black ($\eta=0.0$), blue ($\eta=0.45$), and red ($\eta=0.85$).}
	\label{phir1}
\end{figure}

\begin{figure}[h]
		\subfigure{}{\includegraphics[scale=0.85]{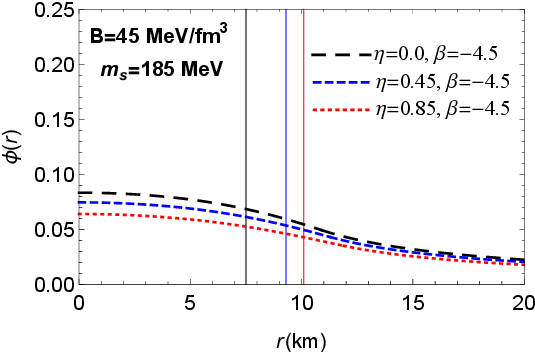}
		}
	\subfigure{}{\includegraphics[scale=0.85]{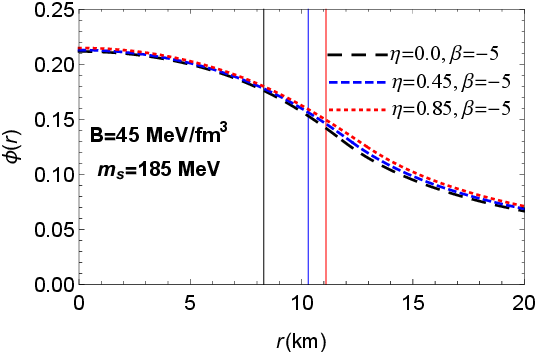}
		}	
\subfigure{}{\includegraphics[scale=0.85]{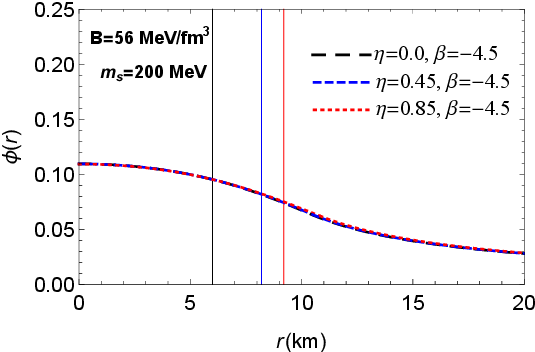}
		}
	\subfigure{}{\includegraphics[scale=0.85]{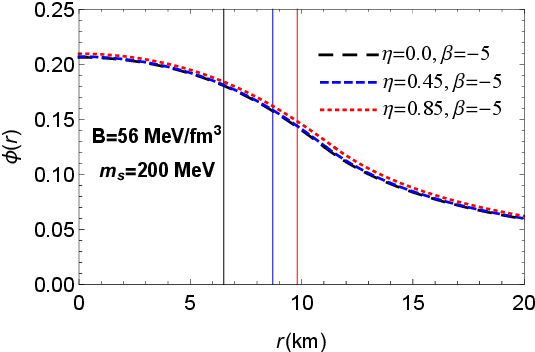}
		}
	\caption{Same as Fig. \ref{phir1} but for the second model of the HNS EoS. We have applied different values of the bag constant, $B$, the mass of the strange quark, $m_s$, and the density jump, $\eta$.}
	\label{phir2}
\end{figure}

\begin{figure}[h]
	\subfigure{}{\includegraphics[scale=0.85]{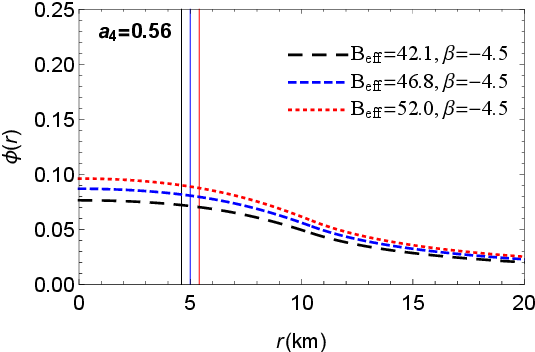}
		}
	\subfigure{}{\includegraphics[scale=0.85]{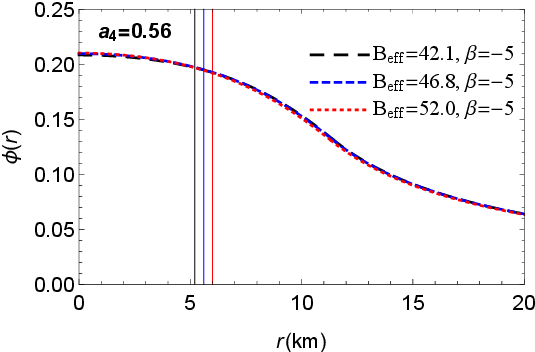}
		}
	\subfigure{}{\includegraphics[scale=0.85]{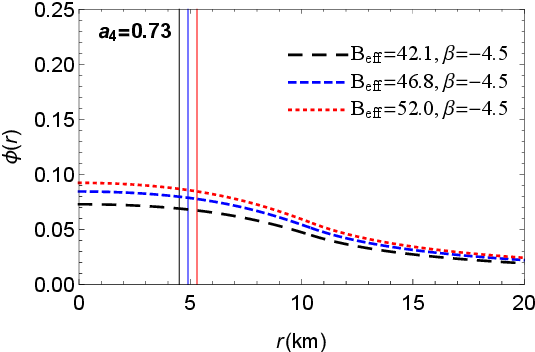}
		}
	\subfigure{}{\includegraphics[scale=0.85]{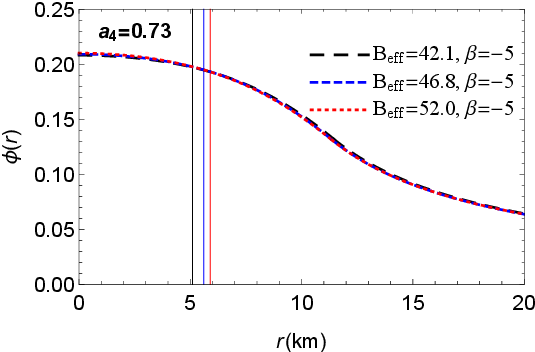}
		}
	\subfigure{}{\includegraphics[scale=0.85]{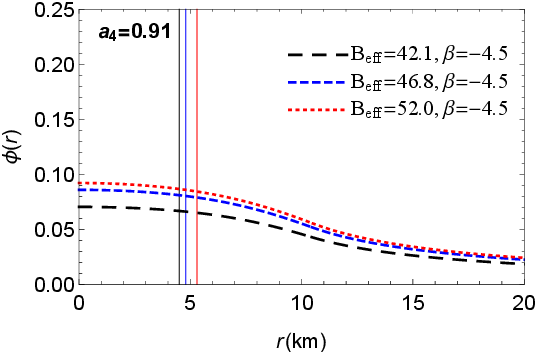}
		}
	\subfigure{}{\includegraphics[scale=0.85]{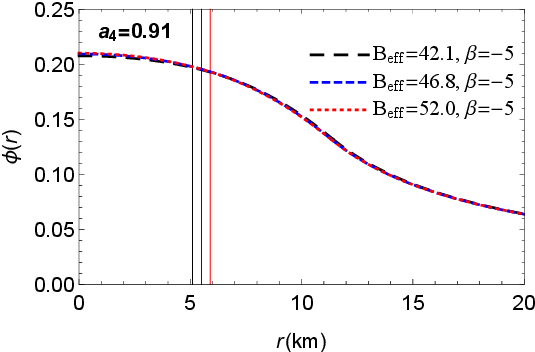}
		}
	\caption{Same as Fig. \ref{phir1} but for the third model of the HNS EoS.
Different values of $B_{eff}$ and $a_4$ with fixed values of the strange quark mass, $m_s = 100\ MeV$, and the density jump, $\eta=0$, have been considered. $B_{eff}$ is in units of $\frac{MeV}{fm^3}$. The position of the phase-splitting surface has been presented by the vertical lines, black ($B_{eff}=42.1$), blue ($B_{eff}=46.8$), and red ($B_{eff}=52.0$).}
	\label{phir3}
\end{figure}

{The profiles of the HNS scalar field considering different models of the HNS EoS have been shown in Figs. \ref{phir1}-\ref{phir3}. In the first model for the HNS EoS, the rate of the scalar field reduction versus the distance from the center of the star is almost independent of the density jump. In this model, the decrease of the coupling constant results in the growth of the scalar field and magnifying the effects of the density jump on the scalar field profile. In the second model of the HNS EoS, the rate of the scalar field reduction can depend on $\eta$, especially in the case with lower values of the bag constant and the strange quark mass (stiffer EoS). Considering the stars with $B=45\frac{MeV}{fm^3}$ and $m_s=185 MeV$, the scalar field as well as its reduction rate decrease by the density jump. In this case, the effects of the density jump on the scalar field profile are more obvious at lower densities. In the third model of the HNS EoS, the growth of the effective bag constant raises the scalar field profile and leads to faster reduction of the scalar field versus the distance from the center of the star.
Fig. \ref{phir3} verifies that the effective bag constant affects $\phi(r)$ more remarkably when the coupling constant is higher. Another impact of the HNS EoS on the scalar field which can be deduced from Fig. \ref{phir3} with $\beta=-4.5$ is that by increasing
the perturbative QCD parameter $a_4$, the scalar field profile reduces. Moreover, the influence of the effective bag constant on $\phi(r)$ is more clear when $a_4$ is larger.}

{We have presented the vertical lines in Figs. \ref{phir1}-\ref{phir3} which denote the position of the phase-splitting surface in each case.
The point that the profile and the corresponding vertical line intersect, gives the value of the scalar field at the phase-splitting surface.
We can find from Figs. \ref{phir1} and \ref{phir2} that the value of the scalar field at the phase-splitting surface is lower when the density jump is larger. Fig. \ref{phir3} proves that in the third model of the HNS EoS with $\beta=-4.5$, the increase in the effective bag constant raises the scalar field at the phase-splitting surface. However, in the case of $\beta=-5$, the value of $\phi(r)$ at the phase-splitting surface decreases by $B_{eff}$. In all cases for the HNS EoS,
the scalar field related to the phase-splitting surface grows by decreasing the coupling constant from $\beta=-4.5$ to $\beta=-5$. In the first and second models, the effects of the density jump on $\phi(r)$ at the phase-splitting surface are more significant with lower values of $\beta$. This is while in the third model, the effective bag constant affects the results more remarkably when $\beta$ is larger. It should be noted that the above results that be realized from Figs. \ref{phir1}-\ref{phir3} have been obtained assuming the stars with $\rho_c=10 \rho_0$. As it is obvious from Figs. \ref{phi1}-\ref{phi3}, in the stars with different central densities, the scalar field behaves differently. }


\subsection{Hybrid neutron star scalar charge}

\begin{figure}[h]
	\subfigure{}{\includegraphics[scale=0.85]{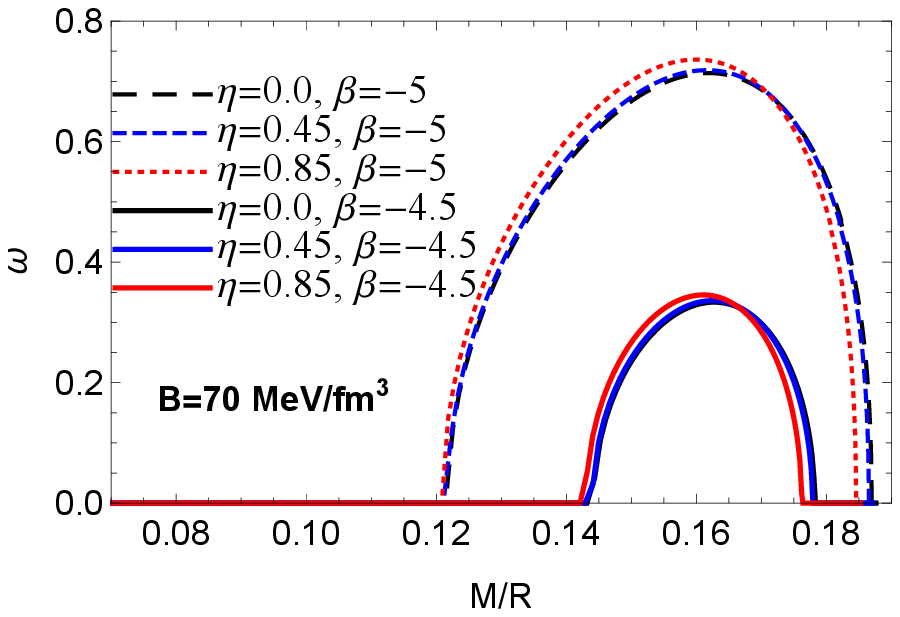}
		}
	\subfigure{}{\includegraphics[scale=0.85]{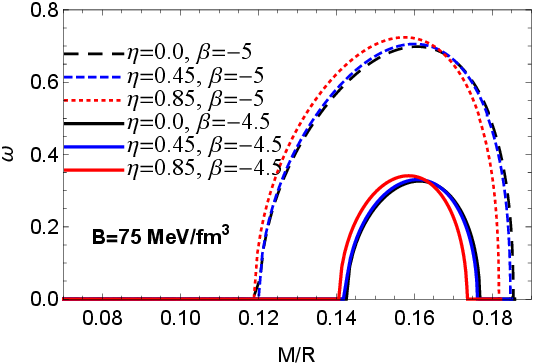}
		}
	\caption{Scalar charge of HNS, $\omega$, versus the star compactness, $M/R$, in the first model of the HNS EoS with different values of the bag constant, $B$, the density jump, $\eta$, and the coupling constant, $\beta$.}
	\label{charge1}
\end{figure}

\begin{figure}[h]
			\subfigure{}{\includegraphics[scale=0.85]{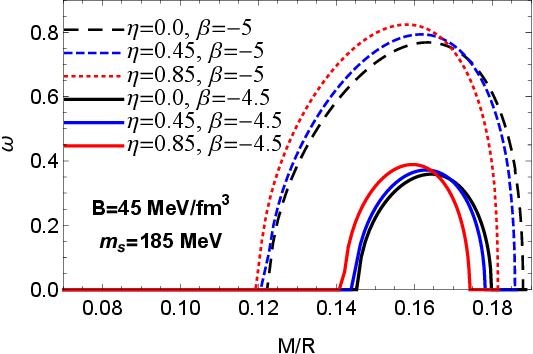}
		}
\subfigure{}{\includegraphics[scale=0.85]{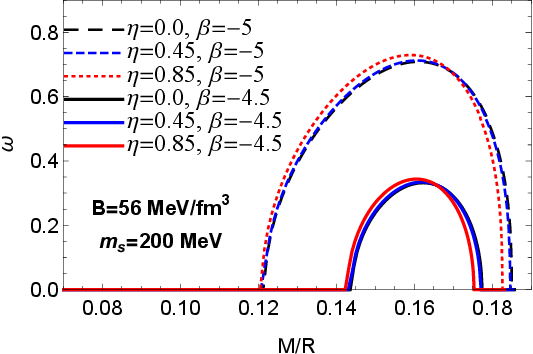}
		}
	\caption{Same as Fig. \ref{charge1} but in the second model of the HNS EoS with different values of the bag constant, $B$, the mass of the strange quark, $m_s$, and the density jump, $\eta$.}
	\label{charge2}
\end{figure}

\begin{figure}[h]
	\subfigure{}{\includegraphics[scale=0.85]{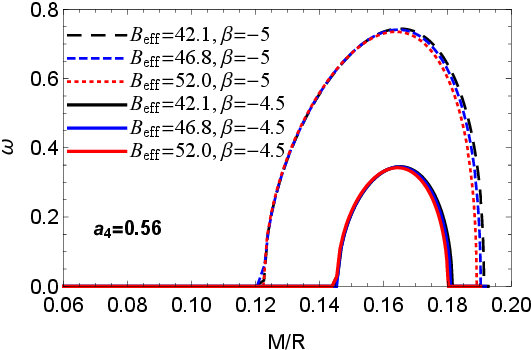}
		}
	\subfigure{}{\includegraphics[scale=0.85]{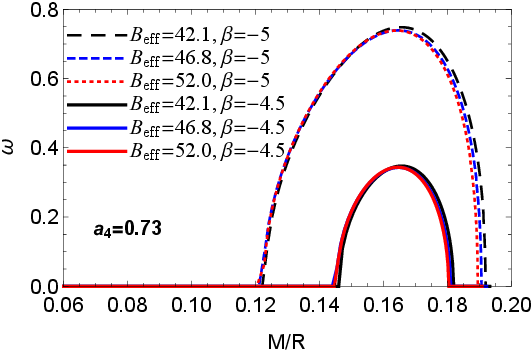}
		}
	\subfigure{}{\includegraphics[scale=0.85]{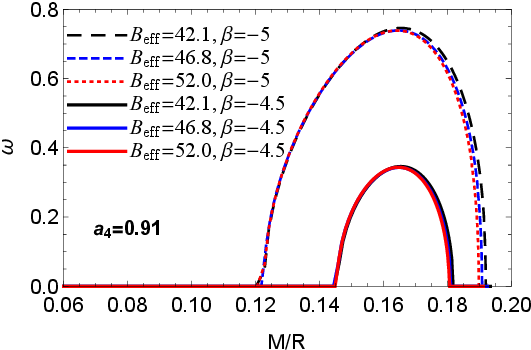}
		}
	\caption{Same as Fig. \ref{charge1} but in the third model of the HNS EoS applying different values of $B_{eff}$ (in units of $\frac{MeV}{fm^3}$) and $a_4$ with fixed values of the strange quark mass, $m_s = 100\ MeV$, and the density jump, $\eta=0$.}
	\label{charge3}
\end{figure}

In Figs. \ref{charge1}-\ref{charge3}, we have plotted the scalar charge of HNS in different models of the HNS EoS.
Scalar charge of scalarized HNS increases by decreasing the coupling constant. With higher values of the density jump {and quark cores with larger density at the phase-splitting surface,} the scalar charge is larger. Considering the stars with larger $\eta$, the less values of the compactness are needed for appearance of the scalar charge (Figs. \ref{charge1} and \ref{charge2}). {In other words, the larger density of the quark matter at the top of the quark core assists the existence of the scalar charge in HNS. } Fig. \ref{charge2} confirms that in the second model of the HNS EoS, the scalar charge decreases when the bag constant and the mass of the strange quark become larger. {This implies that the softer EoS of HNS makes the star scalar charge smaller.} Besides, applying the lower values of the bag constant and the mass of the strange quark, {i.e. stiffer EoS,} the influences of the density jump on the scalar charge are more significant. Fig. \ref{charge3} confirms that with the lower values of the coupling constant, the effective bag constant alters the scalar charge more significantly, {especially in the stars with larger compactness.}

\section{SUMMARY AND CONCLUDING REMARKS}\label{s5}

In the present work, we have studied the hybrid neutron stars (HNSs) in the scalar tensor gravity.
For this aim, a piecewise polytropic EoS constrained by the observational data and different MIT bag models have been employed to describe
the hadronic phase and the strange quark matter, respectively.
{Our calculations approve that the density jump in the HNS and
the higher density of the quark matter at the phase-splitting surface enhance the maximum mass of the scalarized stars.}
{This increase is a result of the stiffening of the HNS EoS due to the density jump.}
{The growth of the bag constant and the mass of the strange quark results in the softening of the HNS EoS and the reduction of he maximum mass of scalarized stars.} Besides, the effective bag constant alters the mass of scalarized stars as well as the central density corresponding to the maximum mass.
{Our calculations verify that the scalarized HNSs in three models are in consistence with different observational constraints.}
In addition, the density jump {and the higher density of the quark matter at the phase-splitting surface} in HNS lead to the reduction of the first and the second critical densities of the spontaneous scalarization. However, the second critical density increases as the effective bag constant grows. The range of the HNS scalarization becomes larger by increasing the effective bag constant {and softening the HNS EoS}. {Besides, the scalar charge of HNS grows as the density jump increases. In fact, the quark cores which have the higher density at the phase-splitting surface enhance the stiffening of the HNS EoS and the scalar charge of stars.}

\acknowledgements{The authors wish to thank the Shiraz University Research Council.}

\end{document}